\documentclass[prx,twocolumn,floatfix,
superscriptaddress,
longbibliography
]{revtex4-1}

\usepackage{amssymb,amsmath,amstext,amsthm,mathtools}
\usepackage{graphicx}
\usepackage{epstopdf}
\usepackage{color}
\usepackage{bm}
\usepackage{appendix}
\usepackage[T1]{fontenc}
\usepackage{bbold}
\usepackage{bbm}
\usepackage{latexsym}
\usepackage[colorlinks=true,citecolor=blue,linkcolor=gray]{hyperref}
\usepackage{lipsum}

\pdfoutput=1
\usepackage[utf8]{inputenc}
\usepackage[english]{babel}
\usepackage[T1]{fontenc}
\usepackage{amsmath}
\usepackage{hyperref}
\usepackage{amsthm}
\usepackage{tikz}
\usetikzlibrary{arrows.meta}
\tikzset{%
  >={Latex[width=2mm,length=2mm]},
            base/.style = {rectangle, rounded corners, draw=black,
                           minimum width=4cm, minimum height=1cm,
                           text centered, font=\sffamily},
  activityStarts/.style = {base, fill=blue!30},
       startstop/.style = {base, fill=red!30},
    activityRuns/.style = {base, fill=green!30},
         process/.style = {base, minimum width=2.5cm, fill=orange!15,
                           font=\ttfamily},
}

\usepackage{gensymb}
\usepackage{hhline}
\usepackage{xcolor}
\usepackage{graphicx}
\usepackage[utf8]{inputenc}
\usepackage[export]{adjustbox}
\usepackage{wrapfig}

\theoremstyle{definition}

\theoremstyle{plain}
\newtheorem{theorem}{Theorem}

\newcommand{\ket}[1]{|#1\rangle}               %ket
\newcommand{\bra}[1]{\langle #1|}              %bra
\usepackage{amssymb}

\usepackage{listings}

\usepackage{stackrel}
\usepackage{atbegshi,picture}

\newcommand{\AC}{\mathcal{A}}

\newcommand{\EC}{\mathcal{E}}

\newcommand{\OC}{\mathcal{O}}

\newcommand{\SC}{\mathcal{S}}

\newcommand{\HS}{\text{HS}}

\newcommand{\Tr}{{\rm Tr}}

\newcommand{\Var}{{\rm Var}}

\renewcommand{\geq}{\geqslant}
\renewcommand{\leq}{\leqslant}

\renewcommand{\vec}[1]{\boldsymbol{#1}}  % Bold vectors instead of arrow vectors

\newcommand{\thv}{\vec{\theta}}
\newcommand{\gamv}{\vec{\gamma}}
\newcommand{\ad}{^\dagger}
\newcommand{\opt}{\rm opt}
\newcommand*{\id}{\openone}
\newcommand{\poly}{\operatorname{poly}}

\newtheorem{proposition}{Proposition}

\newsavebox{\boxA}
\usepackage[ruled,vlined]{algorithm2e}
\include{pythonlisting}
\usepackage{capt-of}

\begin{document}

\title{Variational {H}amiltonian Diagonalization for Dynamical Quantum Simulation}
\date{\today}
\author{Benjamin Commeau}
\affiliation{
Department of Physics, University of Connecticut, Storrs, Connecticut, CT, USA.}
\affiliation{
Information Sciences, Los Alamos National Laboratory, Los Alamos, NM, USA.}
\author{M. Cerezo}
\affiliation{
Theoretical Division, Los Alamos National Laboratory, Los Alamos, NM, USA.}
\affiliation{
Center for Nonlinear Studies, Los Alamos National Laboratory, Los Alamos, NM, USA.}
\author{Zoë Holmes}
\affiliation{
Information Sciences, Los Alamos National Laboratory, Los Alamos, NM, USA.}
\author{Lukasz Cincio}
\affiliation{
Theoretical Division, Los Alamos National Laboratory, Los Alamos, NM, USA.}
\author{Patrick J. Coles}
\affiliation{
Theoretical Division, Los Alamos National Laboratory, Los Alamos, NM, USA.}
\author{Andrew Sornborger}
\affiliation{
Information Sciences, Los Alamos National Laboratory, Los Alamos, NM, USA.}

\begin{abstract}
Dynamical quantum simulation may be one of the first applications to see quantum advantage. However, the circuit depth of standard Trotterization methods can rapidly exceed the coherence time of noisy quantum computers. This has led to recent proposals for variational approaches to dynamical simulation. In this work, we aim to make variational dynamical simulation even more practical and near-term. We propose a new algorithm called Variational Hamiltonian Diagonalization (VHD), which approximately transforms a given Hamiltonian into a diagonal form that can be easily exponentiated. VHD allows for fast forwarding, i.e., simulation beyond the coherence time of the quantum computer with a fixed-depth quantum circuit. It also removes Trotterization error and allows simulation of the entire Hilbert space. We prove an operational meaning for the VHD cost function in terms of the average simulation fidelity. Moreover, we prove that the VHD cost function does not exhibit a shallow-depth barren plateau, i.e., its gradient does not vanish exponentially. Our proof relies on locality of the Hamiltonian, and hence we connect locality to trainability. Our numerical simulations verify that VHD can be used for fast-forwarding dynamics.
\end{abstract}

\maketitle

\section{Introduction}

One of the main motivations for the development of quantum computers has been the potential for simulating quantum systems~\cite{feynman1999simulating}. Quantum algorithms for the simulation of quantum systems have been shown to be exponentially more powerful than corresponding classical algorithms~\cite{lloyd1996universal}. Once realized, quantum simulations are expected to provide transformational advances in the prediction of quantum dynamics with application to quantum foundations~\cite{arrasmith2019variational} and to the design of novel quantum materials~\cite{bauer2020quantum} and, potentially, next-generation quantum computers~\cite{kyaw2020quantum}.

Dynamical simulation algorithms designed for the fault-tolerant era, such as Trotterization methods \cite{lloyd1996universal,sornborger1999higher}, LCU methods \cite{berry2015simulating}, and qubitization methods \cite{low2019hamiltonian}, may lead to prohibitively deep circuits for current quantum computers. In the era of 
Noisy, Intermediate-Scale Quantum (NISQ) devices, Variational Quantum Algorithms~\cite{VQE,mcclean2016theory,qaoa2014,Romero,QAQC,VQSD,arrasmith2019variational,cerezo2020variationalfidelity,bravo-prieto2019,cerezo2020variational,heya2019subspace,cirstoiu2019variational,li2017efficient,endo2018variational,yuan2019theory} provide a promising alternative approach. Variational methods for dynamical simulation break down into conceptually distinct approaches. Extensions of the Variational Quantum Eigensolver have been developed where low-energy subspaces are identified, then integrated in time by scaling their eigenenergies~\cite{heya2019subspace}. Other methods are iterative in time, where state integration is learned with a variational approach, step-by-step~\cite{li2017efficient,endo2018variational,yuan2019theory}. Yet other methods attempt to approximately diagonalize an entire Trotterized unitary, then advance simulation time by modifying eigenenergy-related phases in the diagonalization~\cite{cirstoiu2019variational}.

Simulation methods where a fixed circuit structure is used to integrate quantum dynamics for arbitrary times are called {\it fast-forwarding} methods. For instance, the methods mentioned above for integrating a set of low-lying states~\cite{heya2019subspace} and for integrating an entire Trotterized unitary~\cite{cirstoiu2019variational} are both fast forwarding methods. Fast forwarding is of particular interest in the near term since, if the resulting circuits are of short enough depth, then the circuit can simulate a quantum system for an amount of time determined by errors in the variational algorithm used to derive it. If small enough errors may be achieved, then significant fast forwarding is made possible, allowing for simulation beyond the coherence time of the NISQ device.  While fast-forwarding is not possible for all Hamiltonians~\cite{atia2017fast,BerryEtAl2007}, fast-forwarding is possible for commuting local Hamiltonians \cite{atia2017fast}, quadratic fermionic Hamiltonians \cite{atia2017fast}, continuous-time quantum walks on particular graphs \cite{loke2017efficient}, and the transverse Ising model~\cite{PhysRevA.79.032316}. Hamiltonians that allow for approximate, rather than exact, fast-forwarding are also of significant interest.

The work presented here is designed to make variational quantum simulation both more accurate and more near-term. In particular, we develop a variational algorithm for diagonalizing an entire simulation Hamiltonian. Our algorithm allows for: (1) fast forwarding beyond the coherence time of the quantum computer with a fixed-size quantum circuit, (2) removal of the Trotter error (to the extent possible using an optimization approach), and (3) simulation of an entire system, not just low-lying energy subspaces.

Our proposed algorithm for dynamical simulation is called Variational Hamiltonian Diagonalization (VHD). VHD employs a variational ansatz $W(\theta)$ to approximately transform a given Hamiltonian $H$ into a diagonal form that can be easily exponentiated, e.g., a form composed of local operators. Once this form is found, dynamical simulations for long times can be performed using the same circuit structure as dynamical simulations for short times, simply by changing the time parameter in the exponentiated diagonal form.

We derive two key analytical results related to our variational cost function, which is based on the Hilbert-Schmidt norm and hence is efficiently computable on a quantum device. The first result is that our cost function is operationally meaningful, providing a bound on the average fidelity of the simulation. This operational meaning provides a natural termination condition for the variational portion of VHD. The second result is a theorem that our cost function does not exhibit a shallow-depth barren plateau, i.e., the gradient does not vanish exponentially in the number of qubits. This is a non-trivial result, as it relies on the locality of the Hamiltonian $H$. Moreover, our cost function does not explicitly take the same form as those analyzed in Ref.~\cite{cerezo2020cost}, which studied the gradient scaling for local and global cost functions.

In addition to these analytical results, we present various numerical implementations of VHD, demonstrating that the algorithm works as expected and that it can be used for fast forwarding. We also discuss and implement a method for pre-training our cost function based on unitary diagonalization.

\section{The Variational {H}amiltonian Diagonalization Algorithm}

\begin{figure*}
\begin{center}
\includegraphics[width=1.99\columnwidth]{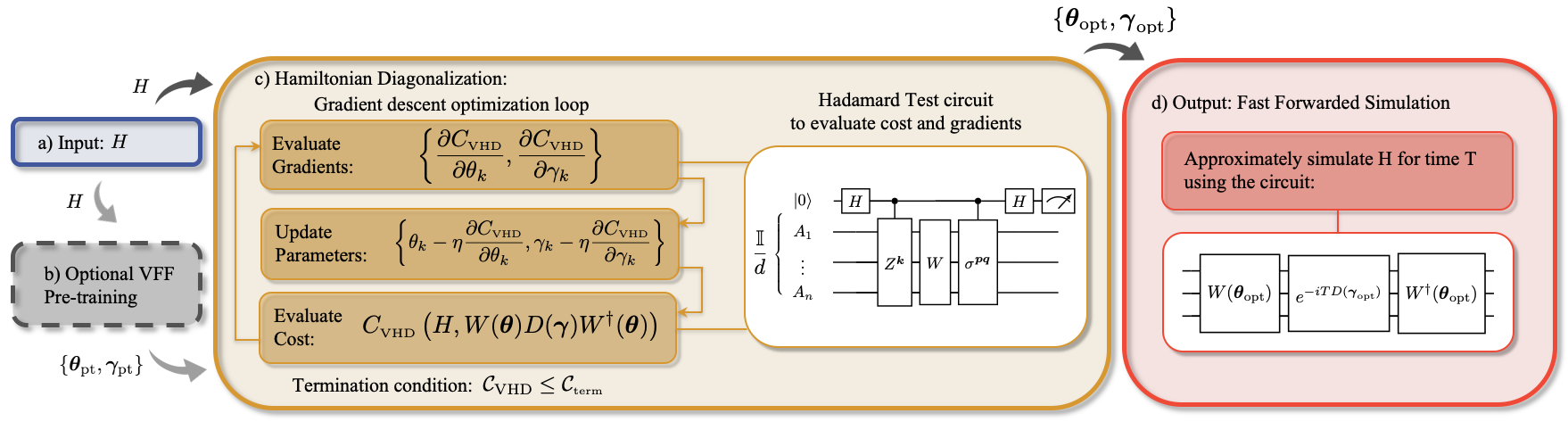}
\makeatletter
\renewcommand{\@makecaption}[2]{%
  \par\vskip\abovecaptionskip\begingroup\small\rmfamily
  \splittopskip=0pt
  \setbox\@tempboxa=\vbox{
    \@arrayparboxrestore \let \\\@normalcr
    \hsize=.5\hsize \advance\hsize-1em
    \let\\\heading@cr
    \@make@capt@title {#1}{#2}
  }
  \vbadness=10000
  \setbox\z@=\vsplit\@tempboxa to .55\ht\@tempboxa
  \setbox\z@=\vtop{\hrule height 0pt \unvbox\z@}
  \setbox\tw@=\vtop{\hrule height 0pt \unvbox\@tempboxa}
  \noindent\box\z@\hfill\box\tw@\par
  \endgroup\vskip \belowcaptionskip
}
\makeatother
\end{center}
\caption{\textbf{The Variational Hamiltonian Diagonalization Algorithm.} (a) The input to the VHD algorithm is a Hamiltonian $H$. (b) Optional pre-training may be performed using the Variational Fast Forwarding (VFF) algorithm of~\cite{cirstoiu2019variational}. (c) The main optimization loop is then used to train the parameters $\{\thv,\gamv\}$ in the ansatz of Eq.~\eqref{eq:VHDansatz}. The Hadamard test circuit shown is used to evaluate the $C_\mathrm{VHD}$ cost terms $c_{\vec{p}\vec{q}\vec{k}}(\thv)=\Tr(\sigma^{\vec{p}\vec{q}}WZ^kW^\dagger)/2^n$, which are real numbers, via the probability of the zero outcome on the ancillary qubit $P(\ket{0})=(1+\Tr(\sigma^{\vec{p}\vec{q}}WZ^kW^\dagger)/2^n)/2$. The optimization loop terminates when the cost function reaches $C_{\text{term}}$ in \eqref{eq:certification}, which guarantees that the simulation will have the desired fidelity. (d) The output from the diagonalization step is then used to implement a fast-forwarded simulation, using Eq.~\eqref{eqn:FFsimulation}.}
\label{fig:flowchart}
\end{figure*}

\subsection{Overview}\label{sec:overview}

The overall structure of the Variational Hamiltonian Diagonalization (VHD) algorithm is shown in Fig.~\ref{fig:flowchart}. The goal of VHD is to diagonalize a target Hamiltonian, and obtain a fixed-structure quantum circuit that approximates the time evolution generated by $H$ up to a time~$T$. 

The input to VHD is a Hamiltonian, $H$, on $n$-qubits (dimension $d = 2^n)$. We assume that $H$ admits an efficient decomposition, with the number of non-trivial non-zero terms being in $\OC(\poly(n))$, in the Pauli basis as 
\begin{align}\label{eq:HPauli}
H=\sum_{\vec{p},\vec{q}}h_{\vec{p}\vec{q}}\sigma^{\vec{p}\vec{q}}\,.
\end{align}
Here, $h_{\vec{p}\vec{q}}$ are real coefficients,  $\sigma^{\vec{p}\vec{q}}=(i)^{\vec{p}\cdot\vec{q}}X^{\vec{p}}Z^{\vec{q}}$ are Pauli strings, and $\vec{p},\vec{q}\in\{0,1\}^{\otimes n}$ are bitstrings of length $n$. In addition, we employ the notation 
\begin{equation}
    X^{\vec{p}}=X_1^{p_1}\otimes\cdots\otimes X_n^{p_n}\,, \quad Z^{\vec{q}}=Z_1^{q_1}\otimes\cdots\otimes Z_n^{q_n}\,,
\end{equation}
with $X_j$ and $Z_j$ Pauli operators acting on qubit $j$. We remark that the decomposition in the Pauli basis in \eqref{eq:HPauli} is taken for simplicity, and in fact the VHD algorithm applies more generally when $H$ can be expressed as a sum of efficiently implementable unitaries (i.e., where the Pauli operators are replaced by more general unitaries).

The first step of VHD is a hybrid quantum-classical optimization loop whose input is $H$ and whose output is a quantum circuit, $W$, and  a diagonal Hamiltonian, $D$, such that $W DW\ad\approx H$. This output can then be used in the second step of VHD, which corresponds to the approximate simulation of the time evolution operator $U(T)=\exp(- i H T)$ as $V(T)=W\exp(-iDT)W\ad$. We now give further details about the individual subroutines in the VHD algorithm.

\subsection{Ansatz}\label{sec:ansatz}

VHD assumes an ansatz for the diagonalization of $H$. This ansatz involves two components: (1) a quantum circuit, $W(\thv)$, that approximately rotates the standard basis into the eigenbasis of $H$, and (2) a diagonal Hamiltonian, $D(\gamv)$, that approximately represents the diagonal form of $H$. Taken together, these two components form the ansatz:  
\begin{align}\label{eq:VHDansatz}
\tilde{H}(\thv,\gamv)= W(\thv)D(\gamv)W^\dagger(\thv) \, ,
\end{align}
whose parameters are trained by the VHD algorithm so that $\tilde{H}$ approximates the target Hamiltonian $H$. We say that VHD can perfectly diagonalize $H$ if there exists a set of parameters $\{\thv,\gamv\}$ such that the training Hamiltonian perfectly matches the target Hamiltonian. In this case, the diagonal elements of $D(\gamv)$ correspond to the eigenvalues of $H$, while  $W(\thv)$ is a matrix whose columns are the eigenvectors of $H$.

Here, the diagonal Hamiltonian operator is given by  
\begin{equation}\label{eq:D}
    D(\gamv) = \sum_{\vec{k}} \gamma_{\vec{k}} Z^{\vec{k}} \, ,
\end{equation}
where $\gamma_{\vec{k}}\in \mathbb{R}$, and where we restrict the number of terms  in~\eqref{eq:D} to be in $\OC(\poly(n))$. While the $Z^{\vec{k}}$ in~\eqref{eq:D} could be general, in practice it may be desirable to assume that these are local operators. Such local operators are easily exponentiated (in the fast-forwarding step of VHD), and also we provide a trainability guarantee (in Theorem~\ref{prop:train} below) in this case. We remark that it has been shown that certain types of Hamiltonians can be diagonalized with purely local terms in the diagonal form~\cite{verstraete2009quantum}, i.e., where the bitstrings $\vec{k}$ are of Hamming weight one.

Regarding the quantum circuit ansatz for $W(\thv)$, keeping the depth short will be important for trainability. This is because deep ansatzes can lead to barren plateaus (i.e., exponentially vanishing gradients), both in the absence~\cite{mcclean2018barren,cerezo2020cost,sharma2020trainability,cerezo2020impact} and presence of noise~\cite{wang2020noise}. Along these lines, it is natural to propose a hardware-efficient ansatz~\cite{kandala2017hardware} for $W(\thv)$. Combining such an ansatz with parameter initialization strategies~\cite{grant2019initialization,volkoff2020large,verdon2019learning} has the potential to mitigate barren plateau issues. Taking this hardware-efficient approach, $W(\thv)$ is expressed as a product of gates from a given alphabet $\AC$ as
\begin{equation}
    W(\thv)=\prod_\mu\exp(- i \theta_\mu G_\mu)W_\mu\,,
\end{equation}
where $G_\mu$ are Hermitian operators, and where $W_\mu$ are unparametrized unitaries. When employing a quantum hardware, $\AC$ is composed of gates native to that specific device. This choice of ansatz reduces the depth overhead when implementing $W(\thv)$.  Specifically, for the numerical implementations in this work we use a layered hardware-efficient ansatz where the gates in $W(\thv)$ act on neighboring qubits in a brick-like structure~\cite{cerezo2020cost}.

\subsection{Cost Function}\label{sec:costfunction}

To quantify how well $\tilde{H}(\thv,\gamv)$ approximates $H$ we define the VHD cost function as the  squared Hilbert-Schmidt distance between the training and target Hamiltonians 
\begin{align}\label{eq:costVHD}
C_{\text{VHD}}(\thv,\gamv) = \frac{||H-\tilde{H}(\thv,\gamv)||_{\HS}^2}{d} \, ,
\end{align}
where $|| X ||_{\HS} = \sqrt{\Tr(X X^\dagger)}$ is the Hilbert-Schmidt norm. Note that this cost function is faithful, vanishing if and only if $\tilde{H}(\thv,\gamv) = H$. Moreover, as shown in Section~\ref{sec:Term}, it is operationally meaningful for non-zero values, with a small cost guaranteeing a large  simulation fidelity. It is convenient to also define a normalized version of the VHD cost as
\begin{align}
\widehat{C}_{\text{VHD}}(\thv,\gamv) = \dfrac{C_{\rm VHD}(\thv,\gamv)}{2\mathcal{N}} \, .
\label{eq:cnhff}
\end{align}
The normalization coefficient $\mathcal{N}=\sum_{\vec{p},\vec{q}}h_{\vec{p}\vec{q}}^2+\sum_{\vec{k}}\gamma_{\vec{k}}^2$ guarantees that $0\leq \widehat{C}_{\text{VHD}}(\thv,\gamv) \leq 1$ irrespective of the system size, and hence is useful for assessing the VHD performance for different problems.

\subsection{Cost Evaluation}\label{sec:costevaluation}

To measure the cost $C_{\text{VHD}}$ it is helpful to expand~\eqref{eq:costVHD} into a summation of terms that can be efficiently evaluated. It is straightforward to verify that the VHD cost function can  be expressed as
\small
\begin{align}
C_{\text{VHD}}(\thv,\gamv) = 
\sum_{\vec{p},\vec{q}}h_{\vec{p}\vec{q}}^2+\sum_{\vec{k}}\gamma_{\vec{k}}^2 -2
\sum_{\vec{p},\vec{q},\vec{k}}h_{\vec{p}\vec{q}}\gamma_{\vec{k}}c_{\vec{p}\vec{q}\vec{k}}(\thv)
\,,
\label{eq:VHD-expansion}
\end{align}
\normalsize
where the $c_{\vec{p}\vec{q}\vec{k}}(\thv)$ are real numbers defined by
\begin{align}\label{eq:cterm}
c_{\vec{p}\vec{q}\vec{k}}(\thv) = 
\dfrac{\Tr\left(\sigma^{\vec{p}\vec{q}}W(\thv)Z^{\vec{k}} W(\thv)^\dagger \right)
}{d} \,.
\end{align}
Note that the first two terms in~\eqref{eq:VHD-expansion} can be classically evaluated. In fact, the first term is fixed, and depends only on $H$. On the other hand, each term $c_{\vec{p}\vec{q}\vec{k}}(\thv)$, and its gradient with respect to any parameter $\theta_\mu\in\thv$, can be efficiently measured using the Hadamard Test depicted in Fig.~\ref{fig:flowchart}(c).  Once all $c_{\vec{p}\vec{q}\vec{k}}(\thv)$ have been measured, the third term of~\eqref{eq:VHD-expansion} can also be classically evaluated, and we remark that this calculation is efficient as there are at most $\OC(\poly(n))$ terms in the summations over $\vec{p},\vec{q}$ and $\vec{k}$. Finally, we note that $\mathcal{N}$, and therefore the normalized cost in~\eqref{eq:cnhff}, can also be efficiently computed.

\subsection{Optimization}\label{sec:optimization}

As shown in Fig.~\ref{fig:flowchart}, the parameters $\{\thv,\gamv\}$ are trained through a hybrid quantum-classical optimization loop. At each iteration step the cost (or its gradient) is estimated for a fixed set of parameters, which are then fed to a classical optimizer that provides updated parameters to solve the optimization problem
\begin{align}\label{eq:optimization}
\{\thv_{\opt},\gamv_{\opt}\}=\underset{\thv,\gamv}{\text{ arg min }} C(\thv,\gamv) \, .
\end{align}
In Section~\ref{sec:Term} we provide an operationally meaningful termination condition for the optimization loop, and in Section~\ref{sec:Train} we analyze conditions under which the trainability of $C_{\text{VHD}}$ is guaranteed.

\medskip

\subsection{Fast-Forwarded Simulation}\label{sec:FFS}

Once the parameters $\thv_{\opt}$ and $\gamv_{\opt}$ that minimize the cost $C_{\text{VHD}}$ have been obtained, VHD employs the Hamiltonian $\tilde{H}(\thv_{\opt}, \gamv_{\opt})$ to approximately simulate the time evolution unitary generated by $H$ up to time $T$. That is, the unitary $U(T)=\exp(- i H T)$ is approximated with 
\begin{align}\label{eqn:FFsimulation}
    V(T)  &= \exp(-i \tilde{H}(\thv_{\opt}, \gamv_{\opt}) T)\nonumber\\
    &= W(\thv_{\opt}) \exp(- i D(\gamv_{\opt}) T ) W(\thv_{\opt})^\dagger \, .
\end{align}
We note that the matrix exponential of $D$ can be exactly implemented since  $[Z^{\vec{k}},Z^{\vec{k}'}]=0$ $\forall \vec{k},\vec{k}'$, meaning that
\begin{equation}\label{eq:expD}
    \exp(- i D(\gamv_{\opt}) T ) = \prod_{\vec{k}}\exp\left(- \gamma_{\vec{k}} T Z^{\vec{k}}\right) \,.
\end{equation}
Here, each unitary $\exp\left(- \gamma_{\vec{k}} T Z^{\vec{k}}\right)$ has an efficient circuit decomposition. For $\vec{k}$ with Hamming weight equal to one, the unitary in~\eqref{eq:expD} is simply given by a tensor product of rotations around the $z$-axis. For larger Hamming weight the exponential of $Z^{\vec{k}}$ can be implemented via CNOT ladders~\cite{welch2014efficient,zhu2018hardware}. In all cases, the circuit depth of $V(T)$ is fixed and does not scale with the length of time simulated, $T$. Therefore VHD potentially allows for long-time simulations at constant depth.
\medskip

\subsection{Optional Pre-training}\label{sec:pretrain}

As discussed in Section~\ref{sec:costevaluation}, evaluating $C_{\rm VHD}$  requires measuring each of the $c_{\vec{p}\vec{q}\vec{k}}(\thv)$ terms separately. There are $N_h N_\gamma$ such terms, where $N_h$ is the number of terms in $H$ (i.e., the number of non-zero $h_{\vec{p}\vec{q}}$ terms) and $N_\gamma $ is the number of non-zero $\gamma_{\vec{k}}$ terms in $D$. As a result, the number of circuits ($ N_h N_\gamma$) that need to be run on the quantum computer is expected to scale polynomially with the system size $n$, and typically this polynomial will be super-linear.

While this scaling is efficient in the system size, one can further reduce the resource requirements of VHD by pre-training the parameters $\{ \thv, \gamv \}$ via the  Variational Fast Forwarding algorithm (VFF) \cite{cirstoiu2019variational}. As shown in~\cite{cirstoiu2019variational, QAQC}, the number of circuits run per optimization step in the VFF algorithm scales linearly with the system size, i.e., linearly in $n$. Thus, this pre-training strategy reduces the total number of calls to the quantum computer during the early stages of the optimization.

While the goal of VHD is to diagonalize the Hamiltonian $H$, the VFF algorithm approximately diagonalizes a Trotterized version~\cite{lloyd1996universal,sornborger1999higher,abrams1997simulation}, $U_\mathrm{TS}(\Delta t)$, of the evolution unitary for small times $U(\Delta t) = \exp(-i H \Delta t)$. 
The approximate diagonalization in VFF is found by compiling $U_{TS}(\Delta t)$ into an operator of the form 
\begin{align}\label{eq:VFFansatz_main_text}
V_{\text{VFF}} = W(\thv_{\text{VFF}}) G(\gamv_{\text{VFF}}) W(\thv_{\text{VFF}})^\dagger \,. 
\end{align}
Here, $W(\thv_{\text{VFF}})$ is a unitary matrix that ideally contains the eigenvectors of $U_{TS}(\Delta t)$ and $G(\gamv_{\text{VFF}})$ is a diagonal operator that ideally represents the eigenvalues of $U_{TS}(\Delta t)$. The compilation is performed using the local Hilbert-Schmidt test~\cite{QAQC} (LHST), which crucially requires measuring only $n$ terms. The LHST circuit is used to evaluate the cost
\small
\begin{align}\label{eq:VFFcostfunction}
  C_{\text{VFF}}(\thv_{\text{VFF}},\gamv_{\text{VFF}}) = 1 -\frac{1}{n} \sum_{j=1}^n F^{e}_j(U_{TS}(\Delta t)\ad V_{\text{VFF}}  )\,,
\end{align}
\normalsize
where $F^{e}_j(U)$ denotes the local entanglement fidelity, with respect to qubit $j$, for the channel defined by the unitary $U$. 
The optimal parameters found from VFF, $\{ \thv_{\rm VFF}, \gamv_{\rm VFF} \}$, determine the pre-trained parameters $\{ \thv_{\rm pt}, \gamv_{\rm pt} \}$ which are used to initialize VHD (See Appendix~\ref{Ap:Pretraining}).

As noted in~\cite{cirstoiu2019variational}, the simulation implemented using VFF necessarily incurs an unavoidable error due to the Trotter-Suzuki approximation. This is not the case for VHD, as the exact Hamiltonian is directly diagonalized. 
In this manner, the combination of pre-training with VFF followed by VHD achieves the best of both worlds: efficient pre-training of the parameters (via VFF), and elimination of Trotter error (via VHD). For more details on the VFF pre-training strategy see Appendix~\ref{Ap:Pretraining}.

\medskip

\section{Theoretical Analysis\label{sec:theory}}

\subsection{Operational meaning and termination condition}\label{sec:Term}

Here we show that the VHD cost function is operationally meaningful, in the sense that small cost values imply high simulation fidelity. Moreover, we  derive a termination condition for the optimization stage of the VHD algorithm.

The success of VHD in approximating the time evolution unitary generated by $H$ can be quantified by the average  fidelity~\cite{nielsen2002simple,horodecki1999general} between $U(T) = \exp(- i H T)$ and $V(T) = \exp(-i\tilde{H} T)$, i.e., by computing 
\begin{align}
\label{eq:AverageFidelity}
\overline{F}(T) =  \int_\psi  d\psi |\bra{\psi} V(T)^\dagger U(T) \ket{\psi}|^2\,.
\end{align}
Here the average is taken over the uniform Haar measure on state space. Then, as explicitly shown in Appendix~\ref{Ap:Term}, the following proposition holds.
\begin{proposition}\label{prop:bound}
Consider the VHD cost functions defined in~\eqref{eq:costVHD}. Then, the following bound holds for all $T$, $\thv$, and $\gamv$:
\begin{align}\label{eq:fidelitybound}
\dfrac{2}{ T^2}
\Bigg(1\!-\!\sqrt{1\!-\!\dfrac{d+1}{d}\Big(1\!-\!\bar{F}(T)\Big)}\Bigg)\!
\leq C_{\text{VHD}}(\thv,\gamv)\, .
\end{align}
\end{proposition}
Equation~\eqref{eq:fidelitybound} confirms that the VHD cost is meaningful with  $\bar{F}(T)=1$ if $C_{\text{VHD}}=\widehat{C}_{\text{VHD}}=0$. Moreover, it establishes the operational meaning of the cost for non-zero values, with a small cost guaranteeing a large final simulation fidelity.

We can further use Eq.~\eqref{eq:fidelitybound} to derive a meaningful termination condition for the VHD optimization loop. Given a desired final simulation fidelity $\bar{F}_{\mbox{\tiny term}}(T)$, we define 
\begin{align}\label{eq:certification}
C_{\mbox{\tiny term}} = \dfrac{2}{ T^2}
\Bigg(1-\sqrt{1-\dfrac{d+1}{d}\Big(1-\bar{F}_{\mbox{\tiny term}}(T)\Big)} \Bigg) \, ,
\end{align}
such that once the cost dips below the termination value $C_{\mbox{\tiny term}}$, i.e., when $C_{\rm VHD}(\thv, \gamv) \leq C_{\mbox{\tiny term}}$, the simulation fidelity is guaranteed to be at least $\bar{F}_{\mbox{\tiny term}}(T)$.  When this condition is satisfied, we terminate the optimization loop and set $\{\thv, \gamv\}=\{\thv_{\opt}, \gamv_{\opt}\}$. 

Finally, we note that given a cost value $C_{\rm VHD}$, one can obtain the guaranteed simulation fidelity by solving for $\bar{F}_{\mbox{\tiny term}}(T)$ in~\eqref{eq:certification}. The latter provides a means of bounding simulation errors.

\subsection{Trainability of the cost}\label{sec:Train}

Recently, it has been shown that variational quantum algorithms and quantum neural networks can exhibit the so-called {\it barren plateau} phenomenon. Here, for random parameter initialization, the gradient and higher order derivatives of the cost function vanish exponentially with $n$~\cite{mcclean2018barren,cerezo2020cost,sharma2020trainability,cerezo2020impact}. For certain cost functions, such barren plateaus even occur when the ansatz is shallow in depth~\cite{cerezo2020cost}. On a barren plateau, exponential precision is required to detect a cost minimizing direction and therefore to navigate through the landscape. Hence, to understand the scaling of VHD, it is paramount to investigate whether $C_{\rm VHD}(\thv,\gamv)$ exhibits a barren plateau. The following theorem, proved in Appendix~\ref{ap:prooftheo}, guarantees that the cost $C_{\rm VHD}(\thv,\gamv)$ does not exhibit a barren plateau under certain conditions.

\begin{theorem}\label{prop:train}
Consider the VHD cost function defined in~\eqref{eq:costVHD}. Then, let $W(\thv)$ be a layered hardware-efficient ansatz with a number of layers in $\OC(\log(n))$, such that each block in the ansatz forms a local $2$-design.  Let the diagonal Hamiltonian $D(\gamv)$ of~\eqref{eq:D} be composed of local operators $Z^{\vec{k}}\in\{Z_j,Z_jZ_{j+1}\}$, and let $H$ be an  $a$-local Hamiltonian with $a\in\OC(\log(n))$. If the coefficients $\gamma_{\vec{k}}$ and $h_{\vec{p}\vec{q}}$ vanish no faster than $\Omega(1/\poly(n))$, then the  variance of the cost function partial derivative $\frac{\partial C_{\rm VHD}(\thv,\gamv)}{\partial \theta_\mu}=\partial_\mu  C_{\rm VHD}$ can be lower bounded as
\begin{equation}\label{eq:VarGradBoundAppendix}
  \Var[\partial_{\mu} C_{\rm VHD}] \geq F(n)  \quad \text{with} \quad  F(n)\in\Omega \left(\frac{1}{\text{poly}(n)} \right)\,.
\end{equation}
Here $\Var[\partial_{\theta_k} C_{\rm VHD}] = \left\langle ( \partial_{\mu} C_{\rm VHD})^2 \right\rangle_{\thv}$, and we used the fact that $\left\langle  \partial_{\mu} C_{\rm VHD}\right\rangle_{\thv}=0$.  Moreover, the expectation value is taken over the angles in $W(\thv)$ for fixed $\gamv$. 
\end{theorem}
Here we recall that an $a$-local Hamiltonian is defined as a Hamiltonian that can be expressed as a sum of terms acting non-trivially on at most $a$ neighboring qubits. 

Let us discuss the implications of Theorem~\ref{prop:train}. First this theorem shows that we can guarantee the trainability of $C_{\rm VHD}$ for Hamiltonians composed of local terms that act non-trivially on less than $\log(n)$ neighboring qubits, and whose associated coefficients are at most polynomially vanishing with $n$. This result establishes a formal connection between locality and trainability. 

Consequently, Theorem~\ref{prop:train} allows us to devise an optimization strategy to avoid barren plateaus. Specifically, the optimization is performed in an inner and an outer loop. In the inner loop, the coefficients $\gamv$ are fixed and one optimizes over $\thv$. From Theorem~\ref{prop:train} we know that there is no barren plateau with respect to $\thv$, and hence, that trainability is guaranteed. In the outer loop, one trains the parameters $\gamv$, and it is clear from inspection of Eq.~\eqref{eq:VHD-expansion} that the gradient with respect to $\gamv$ does not vanish exponentially in $n$. Hence, the trainability of the outer loop is guaranteed from  that of the inner loop.

\section{Numerical Implementations\label{sec:results}}

Here we present results obtained from numerically implementing the VHD algorithm. Specifically, we simulate VHD to  diagonalize and approximate the time evolution generated by a one-dimensional Heisenberg $XY$ model on $n$ qubits
\begin{align}
H_{XY} = \sum_{j=1}^{n-1}
\big(
X_jX_{j+1} + Y_jY_{j+1}
\big)
\,.
\end{align}
Here we recall that the $H_{XY}$ can be exactly diagonalized into a separable Hamiltonian of the form $H = \sum_{j= 1}^n e_j Z_j$ via a Jordan-Wigner transformation~\cite{sachdev2007quantum}. Hence, we take the diagonal Hamiltonian to be composed of local Pauli $Z$ terms, i.e.,  $D = \sum_{j= 1}^n \gamma_j Z_j$.

\begin{figure}[t]
\centering
\includegraphics[width=.95\columnwidth]{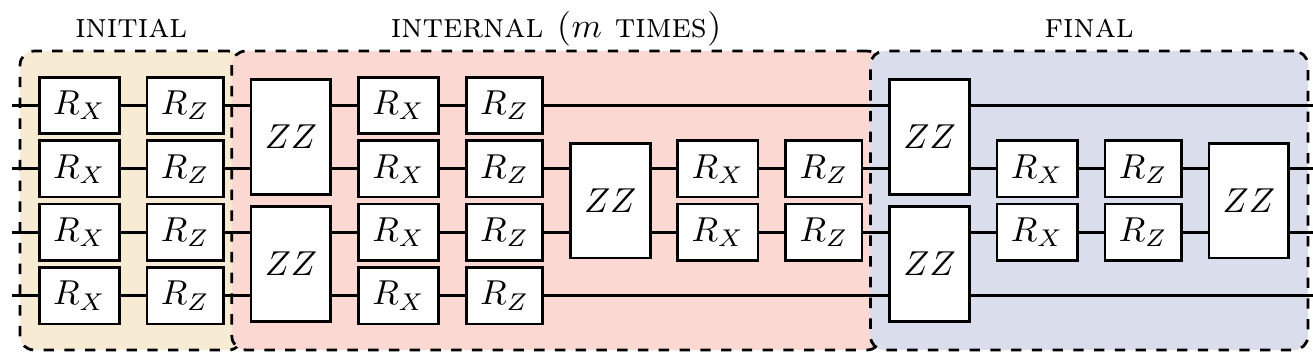}
\caption{\textbf{Layered hardware-efficient ansatz employed in our simulations.} Here  $ZZ=\exp(-i\theta (Z\otimes Z)/2)$ is a two qubit entangling gate, while $R_X$ and $R_Z$ respectively denote rotations around the $x$ and $z$ axes. As depicted, the ansatz is composed of an initial layer of single qubit rotations, followed by $m$ internal layer, and a final layer. Shown is the case of $n=4$ qubits. In our heuristics we have $m=n$. }
\label{fig:numansatz}
\end{figure}

The VHD algorithm was implemented for $n=3,4,5$ qubits, and for each value of $n$ we ran $80$ instances of VHD and we picked the best one. In all cases, we employed a layered hardware-efficient ansatz as depicted in Fig.~\ref{fig:numansatz}. To optimize the parameters, we began with VFF pre-training as described in Sec.~\ref{sec:pretrain}. For this pre-training we employed a first-order Trotter-Suzuki approximation $U_{TS}(\Delta t)$ of the short time evolution of $H_{XY}$ for time $\Delta t = 0.25$. 
After $320$ iteration steps, we switched from pre-training to optimizing the VHD cost of Eq.~\eqref{eq:costVHD} for an additional $320$ iteration steps. The optimization was performed using a gradient descent method, with the gradients being computed via  the parameter shift rule~\cite{mitarai2018quantum,schuld2019evaluating}. The cost and its gradients were evaluated using finite sampling, with $10^9$ shots per cost function evaluation.

\begin{figure}[t]
\centering
\includegraphics[width=0.45\textwidth]{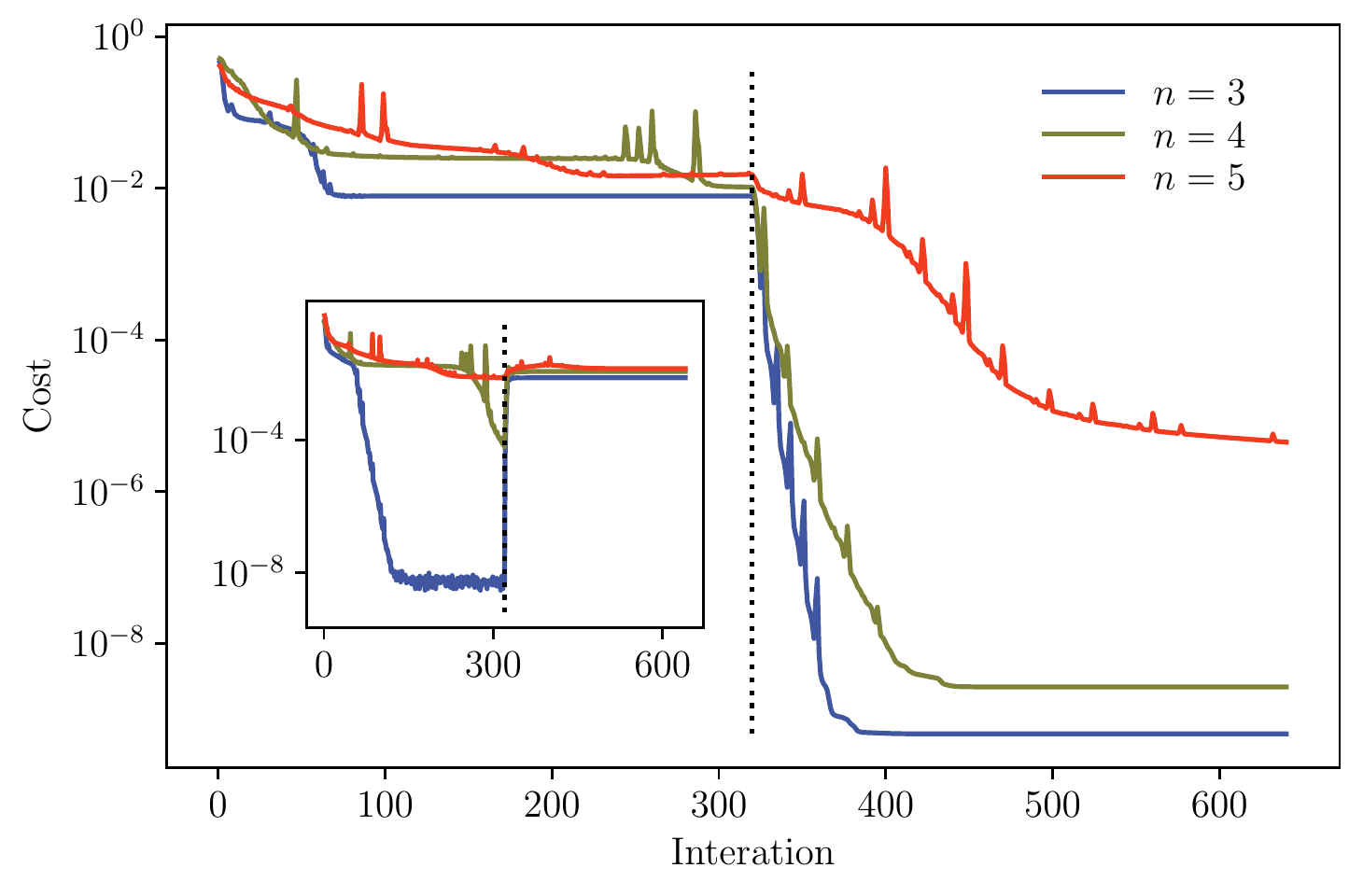}
\caption{\textbf{VHD optimization results with VFF pre-training for the XY Hamiltonian.} The main plot (inset) shows the cost $\widehat{C}_{\mbox{\tiny VHD}}(\thv, \gamv)$ ($C_{\mbox{\tiny VFF}}(\thv, \gamv)$) a function of iteration step for system sizes of 3 (red), 4 (green) and 5 (blue) qubits.  Both the pre-training with VFF and the direct VHD optimization consisted of 320 optimization steps. The black dotted line indicates the end of pre-training.}
\label{fig:Opt}
\end{figure}

Figure~\ref{fig:Opt} shows the VHD cost function versus the number of iterations for the values of $n$ considered. Additionally, the inset depicts the VFF cost function for the same iterations. To compare the performance for different system sizes we plot the normalized cost in \eqref{eq:cnhff}. During pre-training, the VFF cost defined in \eqref{eq:VFFcostfunction} was minimized to $10^{-8}$, $10^{-4}$, and $10^{-3}$ for $n=3,4,$ and $5$, respectively. At each iteration we compute the VHD cost with the parameters trained via VFF, and we see that the cost decreases as the number of iterations increases. This indicates that pre-training allows us to obtain parameters that reduce the VHD cost. However, we can also see that the pre-training does not allow us to fully optimize  $C_{\rm VHD}$, as the VHD  cost function plateaus and cannot keep decreasing. This agrees with the fact that VFF is fundamentally limited by the initial Trotter error of $U_{\rm TS}(\Delta t)$. Once we switch to directly minimizing $C_{\rm VHD}$, the cost function is further reduced by several orders of magnitude as the exact Hamiltonian is diagonalized, thereby eliminating the Trotter error. The cost was successfully reduced below $10^{-9}$, $10^{-8}$, and $10^{-5}$ for the $3$, $4$, and $5$ qubit implementations, respectively.

\begin{figure}
\centering
\includegraphics[width=0.45\textwidth]{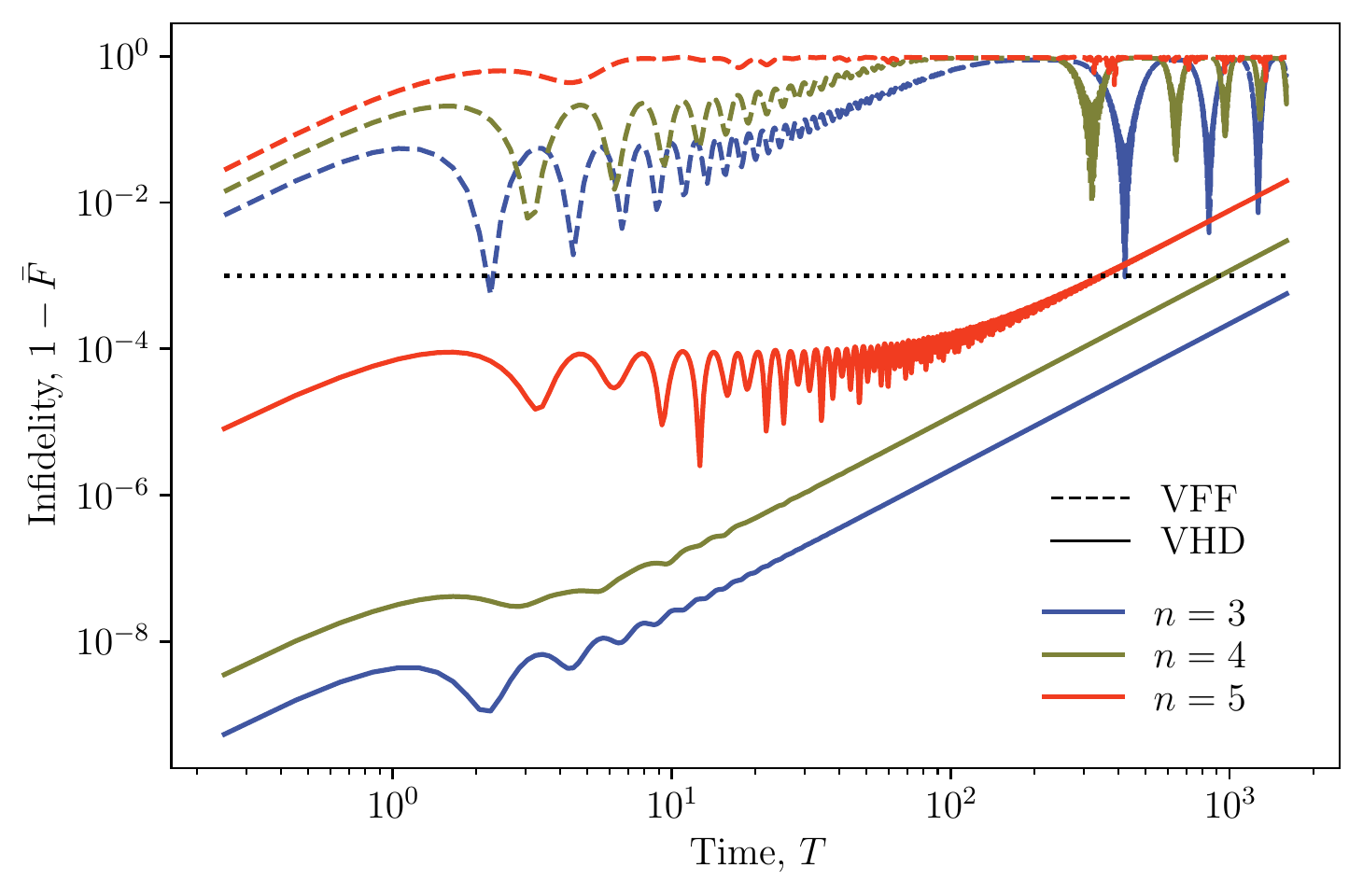}
\caption{\textbf{Fast forwarded quantum simulations of the XY Hamiltonian.} Average simulation infidelity $1 - \bar{F}$ (with  $\bar{F}$ defined in Eq.~\eqref{eq:AverageFidelity}) as a function of time $T$ for systems composed of $n=3$ (red), $n=4$ (green) and $n=5$ (blue) qubits. The fast forwarding is performed using the optimum parameters found after VFF pre-training (dashed lines) and after the full VHD optimization (solid lines) for the optimization runs shown in Fig.~\ref{fig:Opt}. A dotted
black horizontal line is placed at a simulation error tolerance of $10^{-3}$.}
\label{fig:FastForward}
\end{figure}

Figure~\ref{fig:FastForward} plots the fast-forwarded simulation error as a function of time for optimal parameters $\thv_{\rm opt}$ and $\gamv_{\rm opt}$ found after VHD training. As an error measure we use the average simulation infidelity, that is $1 - \bar{F}$ where $\bar{F}$ is defined in Eq.~\eqref{eq:AverageFidelity}. The dashed lines indicate the fast-forwarding that could be achieved using the parameters found from pre-training with VFF and the solid lines indicate the fast-forwarding achieved using the combination of pre-training and then VHD. The VHD algorithm substantially out-performs VFF,
with the simulation infidelity remaining below $10^{-3}$ for times up to $\sim10^3$. In contrast, the infidelity of VFF is greater than $10^{-3}$ by the first time step due to inherent Trotter error. 

\section{Discussion}\label{sec:discussion}

We have introduced a new variational method for quantum simulation that we call Variational Hamiltonian Diagonalization (VHD). This method diagonalizes a quantum Hamiltonian, then via exponentiation, fast forwards the evolution of an initial state. We have demonstrated that our method improves the fidelity of quantum simulations relative to previous variational fast forwarding methods~\cite{cirstoiu2019variational} by removing the Trotterization error from the diagonalization process.

The potential for fast-forwarding with VHD could allow for simulation beyond the coherence time of NISQ computers, unlike standard iterative approaches~\cite{lloyd1996universal,sornborger1999higher,berry2015simulating,low2019hamiltonian} where the circuit depth grows with simulation time. Moroever, VHD is different from other variational simulation algorithms based on optimization over a small set of states ~\cite{li2017efficient,endo2018variational,yuan2019theory,heya2019subspace}, in that it diagonalizes over the entire Hilbert space~\cite{sharma2020reformulation}, which allows for the fast forwarding of higher energy states.

While analytical results for variational quantum algorithms are rare, our work is an exception. Proposition~\ref{prop:bound} gives an operational meaning for the VHD cost function in terms of the average simulation fidelity, which provides a meaningul termination condition for the VHD algorithm. Even more significant is Theorem~\ref{prop:train}, which guarantees the absence of barren plateaus for the VHD cost function under certain conditions. There are only a handful~\cite{mcclean2018barren,cerezo2020cost,sharma2020trainability,cerezo2020impact,volkoff2020large} of analytical gradient scaling results for variational quantum algorihtms. Our work makes an important contribution to this field by establishing that local Hamiltonians can be diagonalized with large cost-function gradients. 

The connection between locality and trainability is highly interesting at the conceptual level. Moreover, it is likely to be practically relevant, e.g., when simulating fermionic systems. Namely, different fermion-to-qubit mappings, such as the Jordan-Wigner and Bravyi-Kitaev mappings, lead to different degrees of Hamiltonian locality~\cite{uvarov2020variational}. Our work suggests that researchers should choose the mapping that leads to the most local Hamiltonian.

\section{Acknowledgements}\label{sec:acknowledgement}

We thank Andrew Arrasmith and Cristina Cirstoiu for helpful discussions. BC and ZH acknowledge support from the Los Alamos National Laboratory (LANL) ASC Beyond Moore's Law project. MC and PJC were supported by the Laboratory Directed Research and Development (LDRD) program of Los Alamos National Laboratory (LANL) under project number 20180628ECR. We acknowledge the LANL LDRD program for support of AS and initial support of BC under project number 20190065DR. MC was also supported by the Center for Nonlinear Studies at LANL. LC was partially supported by LANL LDRD under project number 20200022DR. LC and PJC were also supported by the U.S. Department of Energy (DOE), Office of Science, Basic Energy Sciences, Materials Sciences and Engineering Division, Condensed Matter Theory Program.

\bibliography{mybib.bib}

\appendix

\section{Trainability of the VHD cost function}\label{Ap:Trainability}

In this section we first recall results which allow us to analyze the trainability of the VHD cost. We then provide a proof for Theorem~\ref{prop:train}. 

\subsection{Theoretical Framework}

To establish that $C_{\text{VHD}}$ is trainable, we  first recall the results of~\cite{cerezo2020cost} where it was shown that local costs are trainable when employing a shallow, layered hardware-efficient ansatz. More specifically, the authors considered cost functions of the form 
\begin{equation}\label{eq:GenLocalCost}
    C_{L}(\thv)  = \kappa \Tr[O_{L} \widetilde{V}(\thv) \rho \widetilde{V}(\thv)^\dagger]\,,
\end{equation}
where $\rho$ is a quantum state, and where $O_{L}$ is a local operator which can be expressed as
\begin{equation}\label{eq:LocalOp}
    O_{L} = c_0 \id + \sum_k c_k O_k \, ,
\end{equation}
with $O_k$ acting non-trivially on a system of at most two neighboring qubits.  Moreover,  the ansatz $\widetilde{V}(\thv)$ is a layered hardware-efficient ansatz of $L$ layers,  where $L$ is at most in $O(\log(n))$, such that each block in $\widetilde{V}(\thv)$ forms a local $2$-design. Here we recall that this ansatz is arranged in a brick-like structure of unitary  ``blocks'' acting on neighboring qubits. 

Let us now consider a given parameter $\theta_\mu\in\thv$ belonging to a block $B$ in the $l$-th layer of the ansatz  $\widetilde{V}(\thv)$. Then, let us define $S_{\mu}$ as a subsystem  of  $n_{\mu}=2l$ adjacent qubits such that the gate  $B$ acts on the middle qubits of $S_{\mu}$. Then, for any operator  $O_k$ in~\eqref{eq:LocalOp} acting non-trivially  on the same qubits as $B$, the following bound holds
\begin{equation}\label{eq:lower}
    \Var[\partial_\mu C_L]\geq F_n(l)\,,
\end{equation}
where 
\small
\begin{equation}\label{eq:Fdef}
        F_n(l)=\frac{2^{2l+1}\kappa^2 c_k^2}{5^{L+l+4}}D_{\HS}(\rho^{\mu},\id_{\mu}/d_{\mu})D_{\HS}(O_k,\Tr[O_k]\id/4)\,.
\end{equation}
\normalsize
Here the variance is taken with respect to the parameters in $\widetilde{V}(\thv)$, $D_{\HS}(A,B)=|| A-B ||_{\HS}$ is the Hilbert-Schmidt distance, and $\rho^{\mu}$ is the reduced state of $\rho$ in $S_\mu$. Additionally,  $\id_{\mu}$ denotes the identity on $S_{\mu}$, and $d_{\mu}=2^{n_{\mu}}$ is the dimension of $S_{\mu}$. Note that the lower bound in~\eqref{eq:Fdef} is trivial if  $\rho^{\mu}$ is equal to the identity on $S_\mu$, or if the operator $O_k$ is equal to  $\id$, since in those cases the cost is independent of $\thv$.

Here we remark that Eqs.~\eqref{eq:lower}, and \eqref{eq:Fdef} are obtained from Theorem 2 in~\cite{cerezo2020cost}, which provides a lower bound for $\Var[\partial_\mu C_L]$ in terms of a summation of positive terms, one of which is $F_n(l)$. Moreover, note that if $F_n(l)$ vanishes no faster than $\Omega(1/\poly(n))$, then so does $\Var[\partial_\mu C_L]$.

\subsection{Proof of Theorem~\ref{prop:train}}\label{ap:prooftheo}

Let us now prove Theorem~\ref{prop:train}.

\begin{proof}

Let us analyze the conditions under which the  VHD cost function does not exhibit a barren plateau on the parameters of $W(\thv)$ for fixed coefficients $\gamv$. We  first map the VHD cost function onto a cost of the form of $C_{L}$ in Eq.~\eqref{eq:GenLocalCost}.  Note  that  it is always possible to construct a quantum state from the Hamiltonian $H$ as follows
\begin{equation}
    \rho_H = \frac{H + \lambda_{\rm min} \id}{\Tr[H]+ d \lambda_{\rm min}} \,,
\end{equation}
where $\lambda_{\rm min}$ is the absolute value of the smallest eigenvalue of $H$. Hence, $C_{\text{VHD}}$ can be expressed as 
\small
\begin{equation}
\begin{aligned}
C_{\text{VHD}}(\thv,\gamv) =  &\frac{1}{d}\Big( || H ||^2 + || D (\gamv)||^2 +2\lambda_{\rm min}\Tr[D(\gamv)]\label{eq:cost1}\\
&- 2(\Tr[H]+ d \lambda_{\rm min}) \Tr[D(\gamv)  \widetilde{W}(\thv) \rho_H \widetilde{W}\ad(\thv)  ]\Big)\,,\nonumber
\end{aligned}
\end{equation}
\normalsize
where we used the cyclicity of the trace. Here we defined $\widetilde{W}(\thv)=W\ad(\thv)$, and it is straightforward to see that if $W\ad(\thv)$ is a layered hardware-efficient ansatz where each block forms a $2$-design, then so is $\widetilde{W}(\thv)$. Since the first three terms in~\eqref{eq:cost1} are independent of $\thv$, their partial derivative with respect to any $\theta_\mu\in\thv$ will be zero. Hence, to analyze the trainability of the cost, the relevant part of $C_{\text{VHD}}(\thv,\gamv) $ is
\small
 \begin{align}
\widetilde{C}_{\text{VHD}}(\thv,\gamv)  &= -2\eta \Tr[D(\gamv)  \widetilde{W}(\thv) \rho_H \widetilde{W}\ad(\thv)  ] \,,
\label{eq:HFFBarrenPlateauForm}
\end{align}
\normalsize
where $\eta= (\Tr[H]+ d \lambda_{\rm min})/d$. Equation~\eqref{eq:HFFBarrenPlateauForm} is precisely of the form~\eqref{eq:GenLocalCost}, where $D(\gamv)$ (given by~\eqref{eq:D}) corresponds to the measurement operator. We further remark that $\widetilde{C}_{\text{VHD}}(\thv,\gamv)$ and $C_{\text{VHD}}(\thv,\gamv)$ have the same derivative with respect to any parameter in $\thv$.

Given a parameter $\theta_\mu\in\thv$ belonging to a block $B$ in the $l$-th layer of the ansatz  $\widetilde{W}(\thv)$, we now analyze the scaling of the function $F_n(l)$ of~\eqref{eq:Fdef}.  Hence, we have to compute the Hilbert-Schmidt distances $D_{\HS}(\rho^{\mu}_H,\id_{\mu}/d_{\mu})$ and  $D_{\HS}(Z^{\vec{k}},\Tr[Z^{\vec{k}}]\id/4)$, where $\rho_H^{\mu}$ is the reduced state of $\rho_H$ in subsystem $S_{\mu}$, and where $Z^{\vec{k}}$ is an operator in $D(\gamv)$ acting on the same qubits as $B$ does. Assuming that $D(\gamv)$ is composed of local operators $Z^{\vec{k}}\in\{Z_j,Z_jZ_{j+1}\}$,  we have
\begin{equation}
D_{\HS}(Z^{\vec{k}},\Tr[Z^{\vec{k}}]\id/4)=\Tr\left[(Z^{\vec{k}})^2\right]=4\,.
\end{equation}

Then, let us rewrite the Hamiltonian $H$ as
\begin{equation}
H=\sum_{(\vec{m},\vec{n})\in\SC_{\mu}}h_{\vec{m}\vec{n}} \id_{\overline{\mu}}\otimes \sigma^{\vec{m}\vec{n}}+\sum_{\vec{p}',\vec{q}'}h_{\vec{p}'\vec{q}'} \sigma^{\vec{p}'\vec{q}'}\,,
\end{equation}
where $(\vec{m},\vec{n})\in\SC_{\mu}$ are bitstring of length $n_{\mu}$. Here we define  $\SC_{\mu}$ as the set composed of the bitstrings $(\vec{m},\vec{n})$ whose associated Pauli operators in~\eqref{eq:HPauli}  act non-trivially only on $S_{\mu}$. This allows us to calculate the reduced state
\begin{align}
    \rho_H^{S_{\mu}} = \frac{1}{d_{\mu}\eta}\left(\sum_{(\vec{m},\vec{n})\in\SC_{\mu}}h_{\vec{m}\vec{n}}  \sigma^{\vec{m}\vec{n}} + \lambda_{\rm min} \id_{\mu} \right)\,,
\end{align}
and the Hilbert-Schmidt distance
\begin{align}
    D_{\HS}(\rho^{\mu}_H,\id_{\mu}/d_{\mu})&= \frac{1}{d_{\mu}}\left(\frac{\sum_{(\vec{m},\vec{n})\in\SC_{\mu}}h_{\vec{m}\vec{n}}^2+\lambda_{\rm min}^2}{\eta^2 }-1\right)\label{eq:stateHSD}\,.
\end{align}
Since $H$ is traceless, then~\eqref{eq:stateHSD} simplifies to
\begin{align}
    D_{\HS}(\rho^{\mu}_H,\id_{\mu}/d_{\mu})&= \frac{1}{d_{\mu} }\frac{\sum_{(\vec{m},\vec{n})\in\SC_{\mu}}h_{\vec{m}\vec{n}}^2}{\lambda_{\rm min}^2}\label{eq:stateHSD2}\,.
\end{align}

From~\eqref{eq:stateHSD2} we can see that if $\SC_{\mu}=\emptyset $, i.e., if $H$ is composed of operators which always act non-trivially on more than $2l$ qubits, then we have $D_{\HS}(\rho^{\mu}_H,\id_{\mu}/d_{\mu})=0$ for all $\vec{k}$ and $S_{\mu}$, and hence one cannot guarantee the trainability of the VHD cost. Hence, assuming that $\SC_{\mu}\neq\emptyset$, we find from~\eqref{eq:Fdef} that
\begin{align}
  F_n(l)=\frac{2^{5} \gamma_{\vec{k}}^2}{5^{L+l+4}}\sum_{(\vec{m},\vec{n})\in\SC_{\mu}}h_{\vec{m}\vec{n}}^2\,.
\end{align}
Here we recall that $L+l$ is at most in $\OC(\log(n))$, meaning that  $F(n,\vec{k})$ is in $\Omega(1/\poly(n))$ if $\left(\gamma_{\vec{k}}^2\sum h_{\vec{m}\vec{n}}^2\right)$ is in $\Omega(1/\poly(n))$. 
Hence, if the previous conditions are met, the VHD cost function does not exhibit a barren plateau and we have that
\begin{equation}\label{eq:lowerbound}
     \Var[\partial_\mu C_{\rm VHD}] \geq F_n(l)\,, \quad  \text{with}\quad F_n(l)\in \Omega(1/\poly(n))\,.
\end{equation}

Here we remark that if the Hamiltonian  $H$ is an $a$-local operator with $a\in\OC(\log(n))$, then one can obtain a lower bound of the form~\eqref{eq:lowerbound} for any angle $\theta_\mu$. Hence, defining  $F(n)=\min\{F_n(l)\}$ we obtain ~\eqref{eq:VarGradBoundAppendix} from Theorem~\ref{prop:train}.
\end{proof}

%    |\---/|
%    | ,_, |
%     \_`_/-..----.
%  ___/ `   ' ,""+ \  
% (__...'   __\    |`.___.';
%   (_,...'(_,.`__)/'.....+

Let us further remark that Theorem~\ref{prop:train} also allows us to show that the $\gamv$ parameters of the diagonal matrix are also trainable. Considering the following nested optimization strategy. Let us assume that the $\gamv$ are randomly initialized. Then, for fixed $\gamv$, one trains the unitary $W(\thv)$ in an inner optimization loop. Since the trainability of the parameters $\thv$ is guaranteed from Theorem~\ref{prop:train},  it is possible to minimize $C_{\rm VHD}$ for fixed $\gamv$. Having found this minimum, the $\gamv$ parameters can be optimized in an outer optimization loop. Moreover, one can see that since $C_{\rm VHD}$ is linear in $\gamv$ the gradient $\partial_{\gamma_k} C_{\text{VHD}}$ is simply 
\begin{equation}
\begin{aligned}
\partial_{\gamma_\nu} C_{\text{VHD}} = 
2\gamma_nu - 2\sum_{\vec{p},\vec{q}}h_{\vec{p}\vec{q}}c_{\vec{p}\vec{q}\vec{k}}(\theta)
\,.
\end{aligned}
\end{equation}

Hence, the parameters $\gamv$  are trainable using, for instance, a gradient descent optimization strategy.

\section{Termination Condition for VHD}\label{Ap:Term}

\subsection{Useful Identities}

To derive the termination condition for VHD, we will make use of the following equalities and bounds. 

\medskip

\paragraph*{Schatten $p$-norms and their inequalities.}
We use the standard definition of the Schatten $p$-norms
\begin{align}
||A-B||_p = 
\Bigg(
\Tr\bigg[\Big((A-B)^\dagger(A-B)\Big)^{\frac{p}{2}}\bigg]
\Bigg)^{\frac{1}{p}}
\ ,
\end{align}
where $A$ and $B$ are $d$-dimensional complex matrices. The Hilbert-Schmidt norm is the Schatten norm with $p=2$. 
The Schatten $p$-norms are left $P$ and right $Q$ unitary invariant
\begin{align}
||PAQ||_p=||A||_p
\ 
\label{eq:leftrightunitaryinvariant}
\end{align}
and satisfy the triangle inequality,
\begin{align}
||A-B||_p
\leq
||A||_p+||B||_p \, .
\label{eq:schattenpnormtriangleinequality}
\end{align}

\medskip

\paragraph*{Linear scaling in $N$ inequality.} Let $U$ and $V$ be unitary matrices. Then the following bound holds~\cite{cirstoiu2019variational}
\begin{align}
||U^N-V^N||_p\leq N||U-V||_p
\ ,
\label{eq:Linear Scaling in $N$ Inequality}
\end{align}
for any $N>0$.

\medskip

\paragraph*{Distance bound for linear time evolutions.}

Let $t$ be a real number, and let $A$ and $B$ be hermitian operators. Then, as shown below in Section~\ref{sec:prooflemma}, the following inequality holds
\begin{align}
||e^{itA}-e^{itB}||_p\leq t||A-B||_p \; .
\label{lmm:linear time evolutions inequality}
\end{align}

\medskip

\paragraph*{Phase minimum Hilbert-Schmidt norm inequality.}
The phase minimum Hilbert-Schmidt norm between two unitaries $A$ and $B$ is defined as
\begin{align}
||A-B||_2^{\text{min}}
=
\min_{\phi\in \mathcal{R}} ||A-e^{i\phi}B||_2
\\
=\sqrt{2d-2|\Tr[AB^\dagger]|}
\ ,
\label{eq:phaseminin}
\end{align}
and satisfies the inequality
\begin{align}
||A-B||_2^{\text{min}}
\leq
||A-B||_2
\label{eq:phasemininequality}
\ .
\end{align}

\subsection{Main Proof of Termination Condition}

Proposition~\ref{prop:bound} can now be derived as follows.

\begin{proof}
First, let us recall that $U=\exp(-iHT)$, and $V=\exp(-i\widetilde{H}T)$. 
We start from Eq.~\eqref{eq:phasemininequality}
\begin{align}
||U-V||_2^{\text{min}}
\leq
||U-V||_2
\ ,
\end{align}
and apply Eq.~\eqref{lmm:linear time evolutions inequality} to find
\begin{align}
||U-V||_2^{\text{min}}
\leq
T||H-\tilde{H}||_2
\ .
\label{eq:UNUNlNTHH}
\end{align}

We then use Eq.~\eqref{eq:phaseminin} to express
\begin{align}
||U-V||_2^{\text{min}}
=
\sqrt{2d-2|\Tr[U^\dagger V]|}
\end{align}
which can be rewritten in terms of the average fidelity, Eq.~\eqref{eq:AverageFidelity}, as~\cite{nielsen2002simple} 
\begin{align}
||U-V||_2^{\text{min}} =
\nonumber \\ 
\sqrt{2d}
\sqrt{
1-\sqrt{
1-\dfrac{d+1}{d}\Big(
1-\bar{F}(T)
\Big)
}
}
\ .
\label{eq:UNUNeFUU}
\end{align}
Finally, recalling $C_{\text{VHD}}=||H-\tilde{H}||_2^2/d$, and combining equations~\eqref{eq:UNUNlNTHH} and ~\eqref{eq:UNUNeFUU},  we obtain
\small
\begin{align}
\sqrt{2d}
\sqrt{
1-\sqrt{
1-\dfrac{d+1}{d}\Big(
1-\bar{F}(T)
\Big)
}
}
\leq
T \sqrt{d
C_{\text{VHD}}}
\ ,
\end{align}
\normalsize
which can be rewritten as 
\begin{align}
\dfrac{2}{ T^2}
\Bigg(
1-\sqrt{
1-\dfrac{d+1}{d}\Big(
1-\bar{F}(T)
\Big)
}
\Bigg)
\leq
C_{\text{VHD}}
\,.
\end{align}
\end{proof}

\subsection{Proof of the Distance bound for linear time evolutions}\label{sec:prooflemma}

Here we provide a proof for the ``Distance bound for linear time evolution inequality'', i.e., we prove that 
\begin{align}
||e^{itA}-e^{itB}||_p\leq t||A-B||_p \; .
\label{lmm:linear time evolutions inequality2}
\end{align}

\begin{proof}

We start by applying Eq.~\eqref{eq:leftrightunitaryinvariant}
\begin{align}
||e^{itA}-e^{itB}||_p = ||(e^{itA}e^{-itB}-I)e^{itB}||_p
\\
=||e^{itA}e^{-itB}-I||_p
\ .
\end{align}
Then, let us  express the unitary in integral-derivative form
\begin{align}
e^{itA}e^{-itB}-I
=
\int_{0}^{t}ds\dfrac{\partial}{\partial s}\Big[e^{isA}e^{-isB}\Big]
\\
=
\int_{0}^{t}ds \ 
e^{isA}(iA-iB)e^{-istB}
\ .
\end{align}
Moreover, this integral can be written as the limit of the Riemann sum
\begin{align}
||e^{itA}-e^{itB}||_p =
\bigg|\bigg|
\int_{0}^{t}ds \ 
e^{isA}(iA-iB)e^{-istB}
\bigg|\bigg|_p
\\ =
\bigg|\bigg|
\lim_{\Delta s_{k} \rightarrow 0}
\sum_{s_{k}}^{}\Delta s_{k}  \ 
e^{is_{k}A}(iA-iB)e^{-is_{k}tB}
\bigg|\bigg|_p
\ .
\end{align}
Applying the triangle inequality from Eq.~\eqref{eq:schattenpnormtriangleinequality} with repeated application across the terms of Riemann sum we obtain
\begin{align}
||e^{itA}-e^{itB}||_p 
\leq
\nonumber \\
\lim_{\Delta s_{k} \rightarrow 0}
\sum_{s_{k}}^{}
\bigg|\bigg|
\Delta s_{k}  \ 
e^{is_{k}A}(iA-iB)e^{-is_{k}tB}
\bigg|\bigg|_p
\\=
\lim_{\Delta s_{k} \rightarrow 0}\sum_{s_{k}}^{} 
\Delta s_{k} 
\bigg|\bigg|
e^{is_{k}A}(iA-iB)e^{-is_{k}tB}
\bigg|\bigg|_p
\ .\label{eq:limits}
\end{align}
Then, noting that the limit of the Riemann sum in~\eqref{eq:limits} can be expressed as an integral, we find 
\begin{align}
||e^{itA}-e^{itB}||_p  \leq
\int_{0}^{t}ds
\bigg|\bigg|
e^{isA}(iA-iB)e^{-istB}
\bigg|\bigg|_p
\ .
\end{align}
Finally, applying Eq.~\eqref{eq:leftrightunitaryinvariant} and evaluating the integral leads to
\begin{align}
||e^{itA}-e^{itB}||_p  \leq
\int_{0}^{t}ds
\big|\big|
A-B
\big|\big|_p
\\ =
t
\big|\big|
A-B
\big|\big|_p
\ .
\end{align}
\end{proof}

\section{Details on Pre-training using VFF}\label{Ap:Pretraining}

\begin{figure}
    \centering
\includegraphics[width=0.48\textwidth]{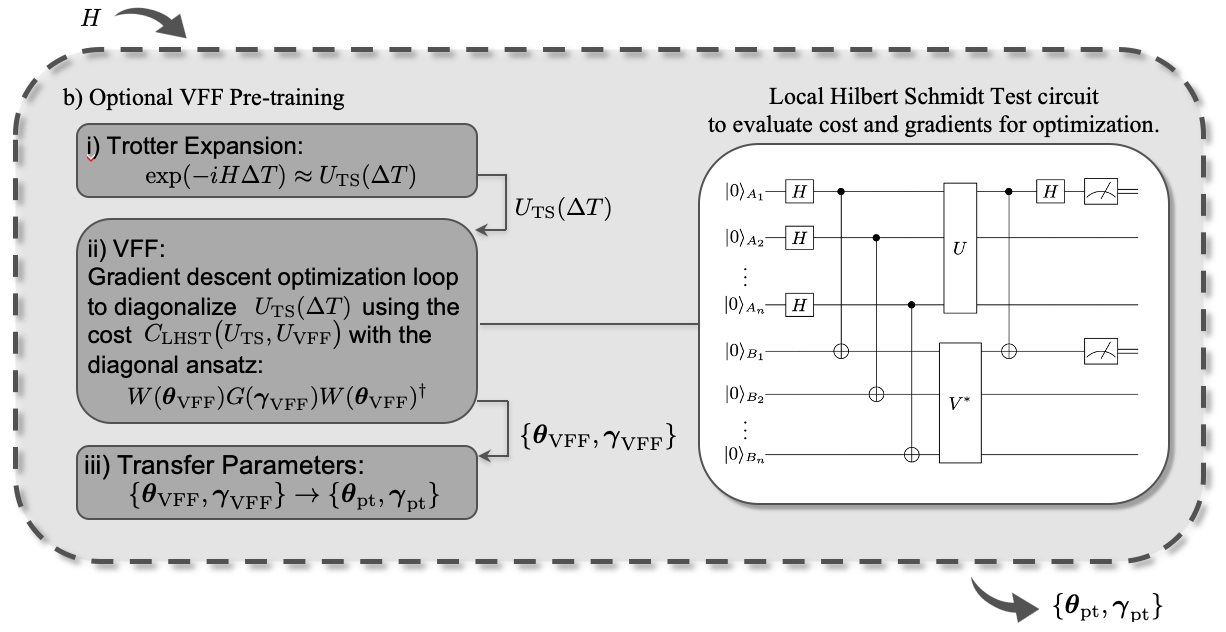}
\caption{\textbf{The VFF Pretraining Algorithm.} The input to VFF pretraining algorithm is a Hamiltonian $H$. (1) The short time evolution induced by $H$, i.e. $U(\Delta t) = \exp(-i H \Delta t)$, is approximated using a Trotter approximation $U_{\rm TS}$. (2) The Trotter unitary $U_{\rm TS}$ is diagonalized using variational compilation using the cost $C_{\rm LHST}(U_{\rm TS}, V_{\rm VFF})$ where $V_{\rm VFF}$ is a diagonal ansatz defined in Eq.~\eqref{eq:DiagonalAnsatz}. The Local Hilbert Schmidt test circuit, shown here, is used to evaluate $C_{\rm LHST}$ with the probability to measure the zero-zero state across the $j_{\rm th}$ pair of qubits giving $F_e^{(j)}$. (3) The output parameters from the diagonalization step $\{\thv_{\rm VFF},\gamv_{\rm VFF}\}$ are finally transformed to the parameters $\{\thv_{\rm pt},\gamv_{\rm pt}\}$ used initialize VHD.}
\label{fig:VFFflow}
\end{figure}

In this section we provide further details on pre-training using VFF. As summarized by the flow chart in Fig.~\ref{fig:VFFflow}, VFF pre-training consists of the following three steps: (1) An initial Trotter approximation $U_{\rm TS}$ of the short time evolution under $H$; (2) A diagonalization of $U_{\rm TS}$ using the local Hilbert-Schmidt test; and (3) A transfer of the optimum parameters found using the local Hilbert-Schmidt test, $\{ \thv_{\rm VFF}, \gamv_{\rm VFF} \}$, to the parameters used to initialize VHD,  $\{ \thv_{\rm pt}, \gamv_{\rm pt} \}$. We detail each of these three steps below.

\subsection{Trotter-Suzuki Approximation of Evolution}\label{sec:tsm}
The first step of VFF pre-training is to approximate the short-time evolution of the Hamiltonian $H$ using a Trotter approximation~\cite{lloyd1996universal,sornborger1999higher,abrams1997simulation}. That is, the unitary evolution for short times, $U(\Delta t) = \exp(-i H \Delta t)$, is approximated via a unitary $U_{\rm TS}$. 

For this paper we approximate the short time evolution using a \textit{first-order} Trotter-Suzuki approximation
\begin{equation}
    U(\Delta t) = U_{\rm TS} + \mathcal{O}\left( (\Delta t)^2 \right)\,,
\end{equation}
with
\begin{align}
U_\text{TS}=\prod_{\vec{p} \vec{q}} e^{-i h_{\vec{p}\vec{q}}\sigma^{\vec{p}\vec{q}} \Delta t}\,, 
\label{eq:st1}
\end{align}
and where $h_{\vec{p} \vec{q}}$ and $\sigma^{\vec{p} \vec{q}}$ are the Pauli coefficients and matrices of $H$, respectively. 

\subsection{Diagonalization using the Local Hilbert-Schmidt Test}\label{ssec:qaqcs}

To approximately diagonalize $U_{\rm TS}$, we variationally compile $U_{\rm TS}$ into ansatz of the form 
\begin{equation}\label{eq:DiagonalAnsatz}
     V_{\rm VFF} := W(\thv_{\rm VFF}) G(\gamv_{\rm VFF}) W(\thv_{\rm VFF})^\dagger \,  .
\end{equation}
Here $W(\thv)$ is a unitary matrix that ideally contains the eigenvectors of $U_{TS}(\Delta t)$ and $G(\gamv_{\rm VFF})$ is a diagonal operator that ideally contains the exponentiated eigenvalues of $U_{TS}(\Delta t)$. 

The compilation is performed by minimizing the cost function $C_{\mbox {\tiny LHST}}$ which is defined as follows~\cite{QAQC}. Let us consider two $n$-qubit registers $A$ and $B$ and let  $A_j$ ($B_j$) represent the $j_{\rm th}$ qubit from the $A$ ($B$) register. We then define
\begin{equation}
    C_{\mbox {\tiny LHST}}(U, V) := 1 -  \frac{1}{n} \sum_{j=1}^n F_e^{(j)} (U, V)\, ,
\end{equation}
where $F_e^{(j)}$ is the entanglement fidelity across the $j_{\rm th}$ pair of qubits. Specifically, the entanglement fidelities $F_e^{(j)}$ are given by
\begin{equation}\label{eq-LHST_prob}
 F_e^{(j)} := 
  \Tr\left(|\Phi^+ \rangle\langle \Phi^+|_{A_jB_j}(\mathcal{E}_{j}\otimes\mathcal{I}_{B_j})(|\Phi^+ \rangle\langle \Phi^+|_{A_jB_j})\right) \, .
\end{equation}
where $\ket{\Phi^+} = (\ket{00} + \ket{11})/\sqrt{2}$ denotes a Bell state and $\EC_j$ is a quantum channel that acts on qubit $A_j$ of the form
\begin{equation}\label{eq-HST_local_channel}
    \EC_j(\rho_{A_j}) = \Tr_{\overline{A}_j}\left(U V\ad \left(\rho_{A_j}\otimes\frac{\id_{\overline{A}_j}}{2^{n-1}}\right)V U\ad \right) \, .
\end{equation}
Here $\overline{A}_j$ is the set of all qubits in $A$ except for $A_j$. 
The fidelities $F_e^{(j)}$ can be evaluated via the Local Hilbert-Schmidt test using the short depth circuit shown in Fig.~\ref{fig:VFFflow}.

The parameters $\{\thv,\gamv\}$ are trained through a hybrid optimization loop with $C_{\rm LHST}$ evaluated on a quantum computer to solve the optimization problem
\begin{align}\label{eq:optimization-appendix}
\{\thv_{\rm VFF},\gamv_{\rm VFF}\}=\underset{\thv,\gamv}{\text{ arg min }} C_{\rm LHST}(\thv,\gamv) \, .
\end{align}
The parameters $\{\thv_{\rm VFF},\gamv_{\rm VFF}\}$ then determine the initial parameters $\{ \thv_{\rm pt} , \gamv_{\rm pt} \}$ for the VHD optimization loop.

\subsection{Transfer of Parameters}

Since $H$ and $\exp(- i H \Delta t)$ share the same eigenvectors, one can use $\thv_{\rm VFF}$ to initialize VHD, that is we can set $\thv_{\rm VFF} = \thv_{\rm pt}$. Note, that since VFF diagonalizes the Trotter unitary $U_{\rm TS}$, and not the exact evolution $\exp(-i H \Delta t)$, the operator $W(\thv_{\rm VFF}) = W(\thv_{\rm pt})$ will of course only capture the eigenvectors of $U_{\rm TS}$ and not $H$ itself.

However, more care needs to be taken relating $\gamv_{\rm VFF}$ and $\gamv_{\rm pt}$, since the diagonal operator $D(\gamv)$ of the VHD ansatz directly captures the eigenvalues of $H$, but the diagonal operator $G(\gamv)$ of the VFF ansatz, captures its exponentiated eigenvalues. We can relate $D$ and $G$ via
\begin{align}
D(\gamv_{\rm pt}) = \frac{1}{-i \Delta t}\ln\left(  G(\gamv_{\rm VFF}) \right) 
\ \, 
\end{align}
which can be rewritten as 
\begin{align}
\sum_k \gamv_{\text{pt}, k} Z^k = \sum_k \ln\left( e^{-i \Delta t \gamv_{\text{VFF}, k} Z^k} \right)
\ \, .
\end{align}
Therefore we have that  
\begin{align}\label{eq:BranchCut}
\gamv_{\rm pt} = \gamv_{\rm VFF} + \dfrac{\pi}{\Delta t}\vec{\alpha}
\ ,
\end{align}
where $\vec{\alpha}$ is a vector of integers. To find $\vec{\alpha}$, we first note that $H$ can be decomposed in terms of the rotated Pauli basis $\{ \tilde{\sigma}_k(\thv) \}$, where $\tilde{\sigma}_k = W(\thv) Z_k W^\dagger(\thv) $, as $H = \sum_k \beta_k(\thv) \tilde{\sigma}_k(\thv)$ where
\begin{equation}
     \beta_k(\thv) = \dfrac{1}{d}\Tr[H \tilde{\sigma}_k (\thv)] =   \sum_{\vec{p}\vec{q}} h_{\vec{p}\vec{q}} c_{\vec{p} \vec{q}k} (\thv)\, . 
\end{equation}
Now, for the transfer of parameters to be effective, we want to choose the vector of integers $\vec{\alpha}$ such that $\gamv_{\rm VFF} + \dfrac{\pi}{\Delta t}\vec{\alpha}$ matches $\vec{\beta}(\thv_{\rm VFF})$ as closely as possible. This is achieved by setting
\begin{align}
   \vec{\alpha} = \text{Round} \left((  \vec{\beta}(\thv_{\rm VFF}) - \gamv_{\rm VFF} ) \dfrac{\Delta t}{\pi} \right) \, 
\end{align}
where the function $\text{Round}(\vec{v})$ finds the nearest integer to each element of $\vec{v}$. Thus we find that $\gamv_{\rm pt}$ and $\gamv_{\rm VFF}$ can be related as 
\begin{equation}
    \gamv_{\rm pt} =  \gamv_{\rm VFF} + \dfrac{\pi}{\Delta t} \text{Round} \left((  \vec{\beta}(\thv_{\rm VFF}) - \gamv_{\rm VFF} ) \dfrac{\Delta t}{\pi} \right) \, .
\end{equation}
The parameters $\{ \thv_{\rm pt}, \gamv_{\rm pt} \}$ can now be used to initialize the VHD optimization algorithm. 
\vfill

\end{document}